\documentclass[preprint,12pt]{elsarticle}
\usepackage{geometry}
\usepackage{amssymb}

\usepackage{graphicx}
\usepackage{amsmath}
\usepackage{enumitem}
\usepackage{float}
\usepackage{xcolor}
\DeclareMathOperator*{\argmax}{arg\,max}
\usepackage[hidelinks]{hyperref}

\usepackage{multirow}
\usepackage{caption}
\usepackage{adjustbox}
\usepackage{tabularx}
\usepackage{subcaption}
\usepackage{longtable}

\journal{Transportation Research Part C}

\begin{document}

\newcommand{\QK}{\color{red}}

\begin{frontmatter}



\title{ICN: Interactive Convolutional Network for Forecasting Travel Demand of Shared Micromobility}


\author[UF]{Yiming Xu\corref{cor1}}

\affiliation[UF]{organization={Department of Civil and Coastal Engineering, University of Florida},
            addressline={1949 Stadium Road}, 
            city={Gainesville},
            postcode={32611}, 
            state={FL},
            country={United States}}

\author[Bloomberg]{Qian Ke}
\author[UF]{Xiaojian Zhang}
\author[UF]{Xilei Zhao}

\affiliation[Bloomberg]{organization={Bloomberg},
            addressline={731 Lexington Avenue}, 
            city={New York City},
            postcode={10022}, 
            state={NY},
            country={United States}}
\cortext[cor1]{Corresponding author. Address: 1949 Stadium Rd, Gainesville, FL 32611. Phone: +01 352-871-3481. Email: yiming.xu@ufl.edu.}

\begin{abstract}
Accurate shared micromobility demand predictions are essential for transportation planning and management. Although deep learning models provide powerful tools to deal with demand prediction problems, studies on forecasting highly-accurate spatiotemporal shared micromobility demand are still lacking. This paper proposes a deep learning model named \textit{Interactive Convolutional Network} (ICN) to forecast spatiotemporal travel demand for shared micromobility. The proposed model develops a novel channel dilation method by utilizing multi-dimensional spatial information (i.e., demographics, functionality, and transportation supply) based on travel behavior knowledge for building the deep learning model. We use the convolution operation to process the dilated tensor to simultaneously capture temporal and spatial dependencies. Based on a binary-tree-structured architecture and interactive convolution, the ICN model extracts features at different temporal resolutions, and then generates predictions using a fully-connected layer. The proposed model is evaluated for two real-world case studies in Chicago, IL, and Austin, TX. The results show that the ICN model significantly outperforms all the selected benchmark models. The model predictions can help the micromobility operators develop optimal vehicle rebalancing schemes and guide cities to better manage the shared micromobility system.
\end{abstract}




\end{frontmatter}


\section{Introduction}
\label{s:1}


Shared micromobility is an innovative transportation strategy that enables users short-term access to a bicycle, scooter, or other low-speed modes on an as-needed basis \cite{shaheen2021shared}. Since the first modern North American bike share system launched in Montreal in 2009, shared micromobility has become a source of innovation, freedom of movement, resilience \cite{nacto2022shared}, and has been experiencing explosive growth in recent years. In 2021, people in the U.S. took 112 million trips on shared micromobility systems, including 62.5 million dockless e-scooter trips, 47 million station-based bike-sharing trips, and 2.5 million dockless bike-sharing trips, contributing to over 70\% increase from 2020 \cite{nacto2022shared}. 

The great growth magnifies some operational challenges faced by shared micromobility systems. The trip origin and destination demands of shared micromobility are spatially and temporally unbalanced \cite{mckenzie2019spatiotemporal, merlin2021segment, xu2020micromobility}. Unbalanced trip demand would lead to the gathering of micromobility vehicles in certain places and spilling out of the parking space, thus blocking sidewalks and causing safety issues. The shared micromibility operators need to rebalance vehicles  to mitigate these adverse effects. The operators collect the redundant vehicles and distribute them to specific locations such as areas where more vehicles are needed. To generate optimal vehicle rebalancing and fleet distribution strategies, unbalanced trip demand needs to be anticipated in advance. Therefore, accurate spatiotemporal shared micromobility demand 
 predictions are essential to efficiently and effectively operate the shared micromobility system \cite{schuijbroek2017inventory, jin2023vehicle}. As shared micromobility is becoming more and more popular in cities and the shared micromobility trips keep increasing, a pressing need exists to study spatiotemporal demand forecasting for shared micromobility services.


Spatiotemporal demand forecasting for shared micromobility can be formulated as a time series prediction problem. Deep learning methods have been proven effective to deal with time series prediction problems due to their flexible model architectures, end-to-end learning processes, and capabilities to capture long-term dependencies and complex non-linear relationships \cite{xu2023real, chen2020predicting, lin2018predicting, li2019learning, pan2019predicting}.
In light of this, this paper proposes a novel deep learning architecture named \textit{Interactive Convolutional Network} (ICN) to deal with the spatiotemporal shared micromobility demand forecasting problem. The proposed model incorporates temporal information (i.e., historical demand and weather) and spatial information (i.e., demographics, functionality, and transportation supply) to achieve a higher prediction performance. We construct extra input data channels by channel dilation operation to embed the spatial information into the input data so that the spatial and temporal dependencies can be learned simultaneously by the convolutional operation. Based on the hierarchical framework and the interactive convolution operations, the ICN model gradually downsamples the dilated input data and extracts features at different temporal resolutions, thus learning an effective representation with enhanced predictability \cite{liu2022scinet}. 

The \textbf{major contributions} of this study are three-fold. First, we developed a new deep learning architecture with an interactive learning mechanism and a novel channel dilation method that accounts for travel behavior knowledge to forecast shared micromobility travel demand. As far as we are aware, the proposed architecture is the first implementation of such an architecture that combines an interactive learning mechanism with spatial information for travel demand forecasting. 
Second, the proposed model demonstrated exceptional forecasting accuracy, significantly outperforming the state-of-the-art benchmark models for the case studies (i.e., Chicago, IL, and Austin, TX).
Third, this study illuminated the under-explored area of real-time travel demand forecasting for an emerging travel mode - shared micromobility. Despite its importance for transportation management, this area, particularly in the context of dockless scooter-sharing, has seldom been investigated.

The remainder of this paper is structured as follows. Section~\ref{s:2} reviews existing literature related to this study. Section~\ref{s:3} formally defines the research problem and describes the proposed method from the overall model architecture to its specific components. Section~\ref{s:4} presents case studies in Chicago, IL, and Austin, TX, and compares the performance of the proposed model with several benchmark models. Section~\ref{s:5} concludes the paper by summarizing findings, identifying limitations, and suggesting future work.

\section{Literature Review}
\label{s:2}
Deep learning methods have provided researchers with powerful tools to deal with forecasting problems in transportation, such as traffic prediction \cite{liu2022scinet, zhao2019t, yu2018spatio, li2018diffusion, guo2019attention, wu2019graph, song2020spatial, bai2020adaptive, jiang2023pdformer}, travel demand forecasting \cite{xu2023real, lin2018predicting, geng2019spatiotemporal, ke2021predicting, liu2019contextualized}, and traffic crash prediction \cite{bao2019spatiotemporal, basso2021deep}. These problems can be formulated as time series modeling tasks, which generate predictions based on the past time steps in the time series. Thus, time series models such as CNN-based models, RNN-based models, graph-based models, and attention-based models have been widely used in these applications.

Traffic prediction is a task that forecasts traffic conditions, such as the traffic volume and speed, at certain spots in a road network. It is one of the most popular forecasting problems in transportation and has been intensively studied by researchers in the past several decades. In this task, the fundamental challenge is to effectively capture the spatial-temporal traffic dynamics \cite{yin2021deep, jiang2023pdformer}. In the past decades, the forecasting model evolved from temporal models and spatial models \cite{zhang2017deep, fu2016using} to integrated spatiotemporal models that capture spatial and temporal dependency simultaneously, such as T-GCN \cite{zhao2019t}, STGCN \cite{yu2018spatio}, DCRNN, \cite{li2018diffusion}, GraphWaveNet \cite{wu2019graph}, ASTGCN \cite{guo2019attention}, AGCRN \cite{bai2020adaptive}, STMGT \cite{xu2023real}, and so on. These models used convolutional neural networks (CNNs) and graph convolutional networks (GCNs) to process spatial representations (e.g., graphs), and used recurrent neural networks (RNNs) and attention-based models (e.g., Transformer) to capture temporal dependency. Regarding the propagation mechanism of traffic flow, these models assumed that traffic speed and volume are highly related in adjacent spots in a road network. Thus, the spatial representations in these models were usually based on spatial adjacency.
Compared with traffic prediction tasks, travel demand forecasting is more challenging. Travel demand is associated with zonal characteristics such as socioeconomic, demographic, and built environment factors rather than zonal adjacency. These contexts should be included in the model for better performance. 

Recently, deep learning models have been widely used in travel demand prediction problems, such as ride-hailing demand prediction \cite{geng2019spatiotemporal}, ridesourcing demand prediction \cite{ke2021predicting}, taxi demand prediction \cite{liu2019contextualized, xu2017real, yao2018deep}, and bike-sharing demand prediction \cite{lin2018predicting, li2019learning, kim2019graph}. These studies indicated that deep learning methods outperformed the classical machine learning models (e.g., random forest and gradient boosting decision tree) and the statistical models (e.g., linear regression and ARIMA). The convolutional neural network (CNN) based models, the recurrent neural network (RNN) based models such as gated recurrent unit (GRU) and long short-term memory (LSTM), and the attention based models such as Transformer are usually used to capture the temporal dependencies \cite{chen2020predicting, pan2019predicting, li2021intercity}. Some studies also combined different neural networks into a spatiotemporal model to capture spatial and temporal dependency simultaneously, and the results showed that the spatiotemporal models have better performance than the single CNN or RNN models \cite{tang2021multi, geng2019spatiotemporal, yao2018deep, li2021intercity}. However, these spatiotemporal models suffer from time-consuming training processes and limited scalability for modeling long sequences due to model complexity \cite{cai2020traffic, xu2020spatial}.

Different kinds of deep learning models have been applied for shared micromobility demand prediction in existing studies, including the RNN models \cite{chen2020predicting, pan2019predicting} and the spatiotemporal models \cite{lin2018predicting, liu2019contextualized, xu2023real}. However, some spatial factors such as the zonal functionality, demographic, and built environment characteristics (e.g., population density) that are highly related to passenger trip demand \cite{bai2020dockless, xu2021identifying} are usually omitted in these models. Although certain models (e.g., \cite{xu2023real}) captured comprehensive spatial and temporal dependencies simultaneously using state-of-the-art time series modeling methods and graph representations, they suffer from intense increases in computational complexity as the graph size grows. 
In addition, despite some recent efforts on the dockless shared micromobility demand forecasting \cite{xu2023real}, most of these shared micromobility demand forecasting studies focused on station-based (i.e., docked) bike-sharing. Studies on real-time spatiotemporal demand forecasting for dockless shared micromobility such as dockless bike-sharing and dockless scooter-sharing are still lacking.

\section{Methodology}
\label{s:3}
In this section, we formally define the research problem, describe the overall architecture of the proposed ICN model, and introduce the detailed model components, including the channel dilation method that enables spatial information embedding, the convolution module, the interactive convolution (IC) block that downsamples input data and conducts interactive learning, and the interactive convolution (IC) network that stacks IC block hierarchically to generate the prediction.  

\subsection{Research problem definition}
We first formally define the shared micromobility demand prediction problem. Given a historical demand matrix $\textbf{X}_t \in \mathbb{R}^{N \times T}$ and other features represented by $\textbf{Z}$, learn a function $f:\mathbb{R}^{N \times T} \rightarrow \mathbb{R}^{N\times M}$ that maps historical shared micromobility demand $\textbf{X}_t$ to the demand in next $M$ time intervals $\textbf{Y}_t \in \mathbb{R}^{N \times M}$:
\begin{equation}
\textbf{Y}_t = [X_{t+1},\dots,X_{t+M}] = f(\textbf{X}_t,\textbf{Z})
\end{equation}
where $\textbf{X}_t = [X_{t-T+1},\dots,X_t]$ is the historical demand matrix at time $t$, $T$ is the  sequence length of input historical demand (i.e., look-back window length), $X_t = [x_t^1,x_t^2,...,x_t^N]$ is the demand at time $t$, $x_t^i$ is the demand of area $i$ at time $t$, $N$ is the total number of areas, $M$ is the length of forecasting time sequence (i.e., forecast horizon).

\subsection{Overview of model framework}
We propose a \textit{Interactive Convolutional Network} (ICN) model to solve the research problem. The proposed model is inspired by the SCINet \citep{liu2022scinet}, which is designed for time series forecasting tasks and has achieved significant forecasting accuracy improvements over existing convolutional models and Transformer-based solutions in traffic prediction tasks \citep{liu2022scinet}. In addition, as a CNN-based model, SCINet is computationally efficient compared with Transformer-based models thus is appropriate for real-time demand forecasting problems. 
Compared with the SCINet \cite{liu2022scinet}, the proposed ICN model incorporates spatial information, including demographics, functionality, and transportation supply, by constructing extra input data channels (i.e., channel dilation). The process of incorporating spatial information will be introduced in Section~\ref{sec:chdlt}. As weather information has a significant impact on the usage of micromobility \cite{xu2023real, noland2021scootin}, we also enable the proposed ICN model to include weather information. 

\begin{figure}[H]
    \centering
    \includegraphics[width=\textwidth]{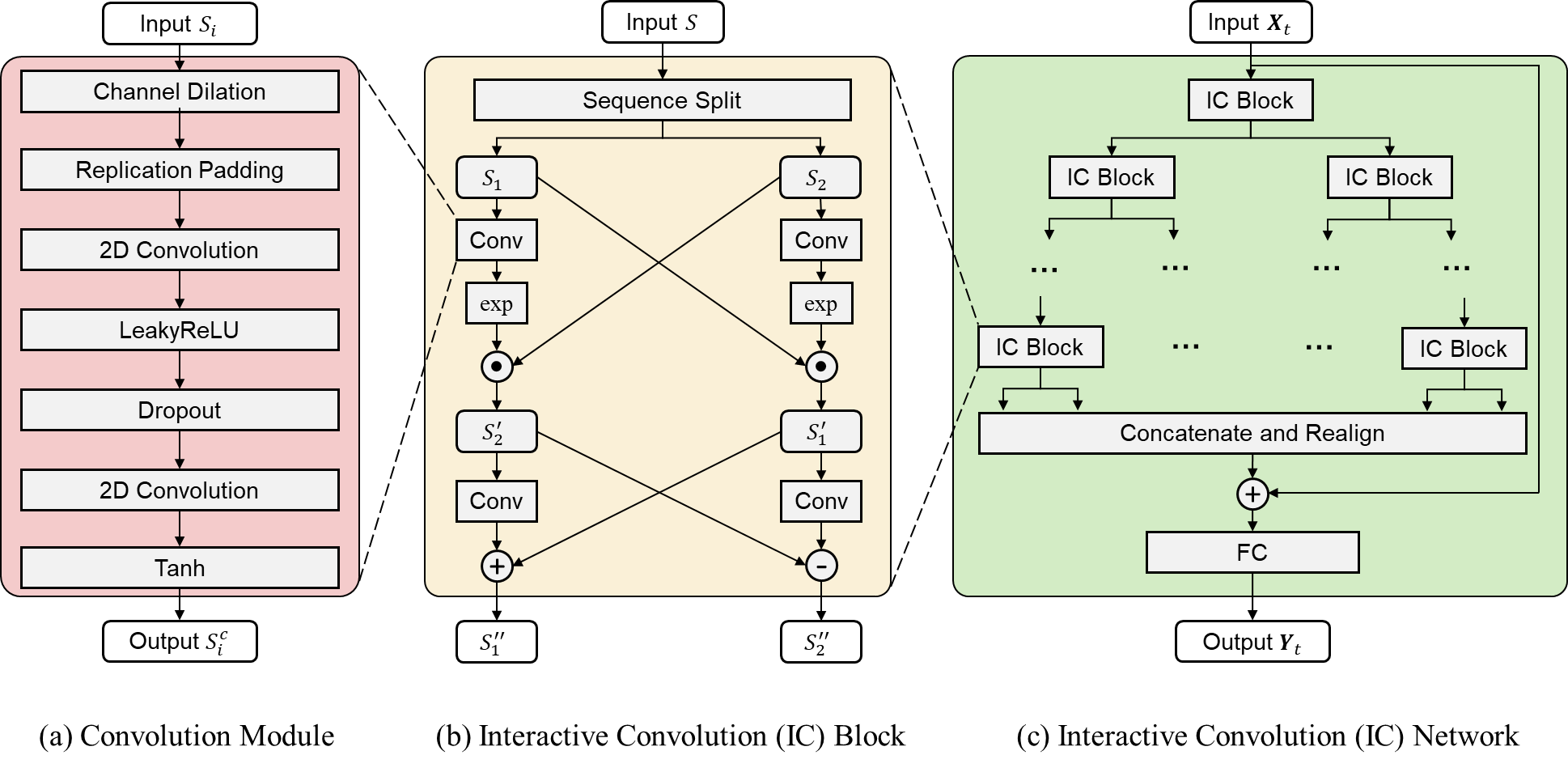}
    \caption{Overall model architecture. (a) Convolution module is a stack of neural network layers. It embeds spatial information using channel dilation and extracts features using convolution operations. (b) Interactive Convolution (IC) Block downsamples the input data and uses convolution operations and interactive learning to extract features at different temporal resolutions. (c) Interactive Convolution (IC) Network stacks IC blocks in a binary-tree structure then uses a fully-connected layer to generate the prediction.}
    \label{fig:frwk}
\end{figure}

The overall framework of the proposed ICN model is presented in Figure~\ref{fig:frwk}. As shown in Figure~\ref{fig:frwk}(a), the convolution module processes the input using a set of operations, including channel dilation, padding, convolution, activation, and dropout. The IC block (Figure~\ref{fig:frwk}(b)) splits the input sequence into two subsequences, then processes the subsequences using convolution modules and interactive learning mechanism. The IC block is stacked to construct a binary-tree-structured IC network (Figure~\ref{fig:frwk}(c)). The details of the convolution module, IC block, and IC network will be introduced in the following subsections.

\subsection{Channel Dilation}
\label{sec:chdlt}

Learning spatial correlations (e.g., characteristic similarity and spatial adjacency) between different areas can enhance the model performance for travel demand prediction \cite{tang2021multi, ke2021predicting}. For example, the areas with similar demographic characteristics tend to have similar travel demand patterns \cite{bai2020dockless, merlin2021segment}, and that information can be used in prediction models to improve the model performance. In the proposed ICN model, we use channel dilation to embed the spatial information into the input time sequence data and use convolutional operations to capture spatial and temporal dependencies simultaneously. To do this, we create extra data channels and concatenate them to the origin input channel $C_0$. According to previous studies, the shared micromobility demand is related to spatial factors such as demographics, land use, and built environment \cite{xu2023real, bai2020dockless, merlin2021segment}. Therefore, we incorporate demographics, functionality (can represent land use features), and transportation supply (i.e., transportation-related built environment features) information by creating three extra channels (i.e., demographic channel, functionality channel, and transportation supply channel), denoted as $C_d$, $C_f$ and $C_t$  correspondingly. The basic idea behind creating an extra channel is to incorporate historical demand information from a correlated area to current area, which helps capture spatial relationships and patterns that can improve the accuracy of demand forecasting.

Let $C_0^i$ denote the historical demand sequence of area $i$ in origin input channel $C_0$ (i.e., $i$th row in $C_0$), the corresponding row $C_d^i$ in demographic channel $C_d$ is given by:

\begin{align}
    C_d^i &= C_0^{j_d}\\
   \textbf{s.t.} \quad j_d &= \argmax_{j\in [0,N],j\neq i}Corr(p_i,p_j)
\end{align}
where $p_i \in \mathbb{R}^{1 \times n_d}$ and $p_j\in \mathbb{R}^{1 \times n_d}$ are the vectors of demographic features of area $i$ and area $j$ respectively, $n_d$ is the number of demographic features, $N$ is total number of areas, $Corr(\cdot)$ is the Pearson correlation coefficient between two areas. 

Similarly, the corresponding row $C_f^i$ in functionality channel $C_f$ and the corresponding row $C_t^i$ in transportation supply channel $C_t$ can be calculated by:

\begin{align}
    C_f^i &= C_0^{j_f}\\
    \textbf{s.t.} \quad j_f &= \argmax_{j\in [0,N],j\neq i}Corr(q_i,q_j)\\
    \nonumber \\  
    C_t^i &= C_0^{j_t}\\
    \textbf{s.t.} \quad j_t &= \argmax_{j\in [0,N],j\neq i}Corr(w_i,w_j)
\end{align}
where $q_i \in \mathbb{R}^{1 \times n_f}$ and $q_j\in \mathbb{R}^{1 \times n_f}$ are the vectors of functionality features of area $i$ and area $j$ respectively, $n_f$ is the number of functionality features, $w_i \in \mathbb{R}^{1 \times n_t}$ and $w_j\in \mathbb{R}^{1 \times n_t}$ are the vectors of transportation supply features of area $i$ and area $j$ respectively, $n_t$ is the number of transportation supply features. To be noted, in all experiment setups, we only consider static features for all three groups of features.

By channel dilation, the model embeds multi-dimensional spatial information adapted from domain knowledge into the input time sequence data to achieve better model performance. Behavioral theories and empirical findings have shown that the shared micromobility travel demand is associated with spatial dependencies (e.g., population density and income) \cite{bai2020dockless, merlin2021segment}. Therefore, neighborhoods with similar spatial dependencies are more likely to have similar travel demand trends. Besides the historical demand in the target area, the historical demand in areas with similar spatial dependencies can also reveal some demand patterns in the target area and provide valuable information for demand forecasting. The channel dilation process embeds this information by creating extra channels and uses convolution operations to capture the embedded spatial information and the temporal dependencies simultaneously. This process enables the spatial information to effectively contribute to forecasting, thus enhancing the model performance.

\subsection{Convolution Module}
\label{sec:conv_module}
\begin{figure}[H]
    \centering
    \includegraphics[width=\textwidth]{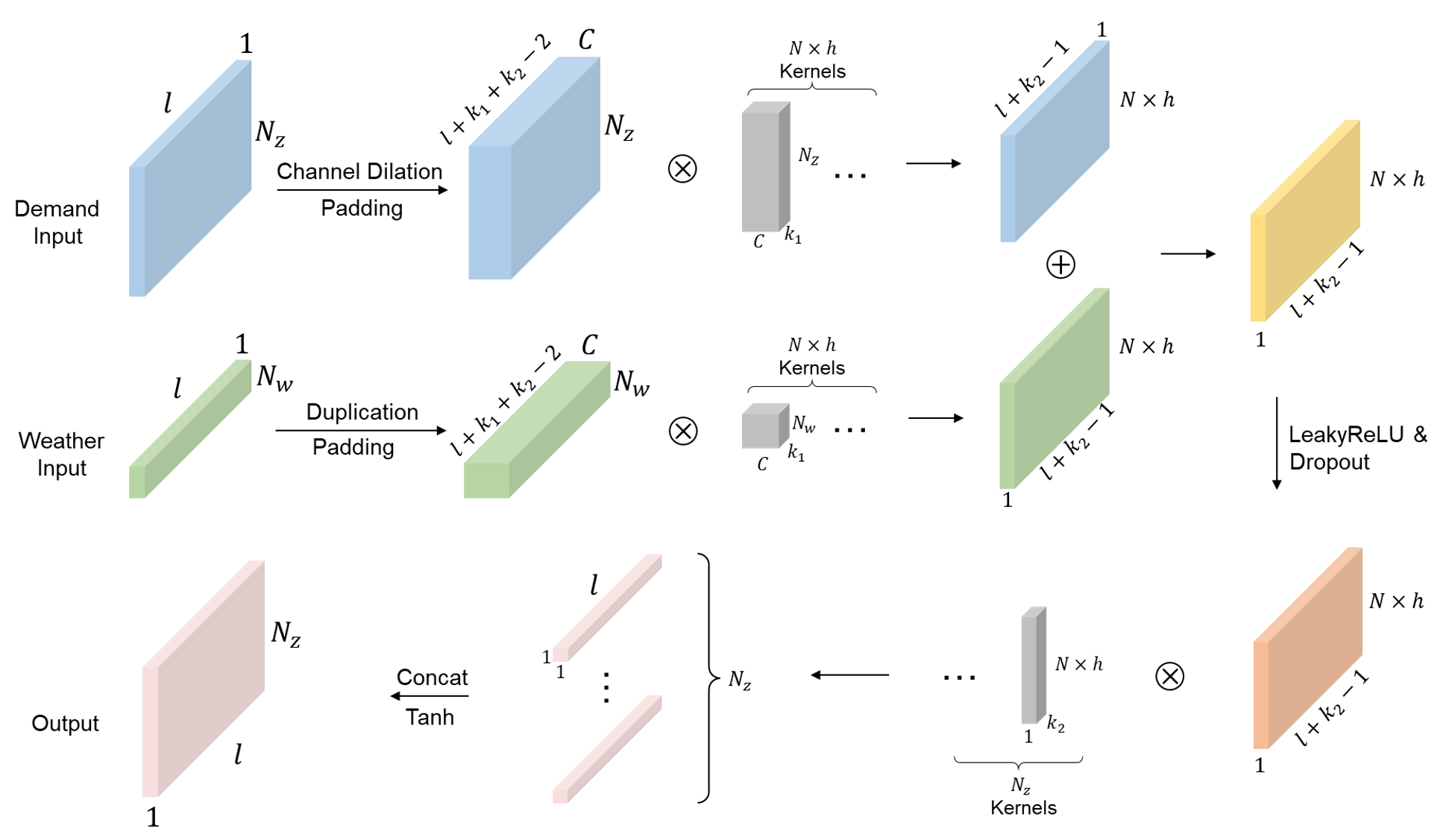}
    \caption{Convolution module}
    \label{fig:conv2d}
\end{figure}

As the spatial information is embedded in the input sequence by the channel dilation process, we use convolution operations to extract spatial and temporal dependencies simultaneously. The channel dilation and convolution operations are conducted by the convolution module, a stack of several basic neural network layers (Figure~\ref{fig:conv2d}). 

The convolution module first dilates the demand input to $C$ channels based on spatial information, then uses a padding layer to pad the input tensor by replicating the boundary. The padded tensor is then processed by a 2D convolution layer with $H = N \times h$ kernels, each kernel with size $[N_Z,k_1]$ and stride 1. At the same time, the weather input is duplicated to $C$ channels, padded by a replication padding layer, and processed by a 2D convolution layer with $H = N \times h$ kernels, each kernel with size $[N_w,k_1]$ and stride 1. Then we add the two convoluted tensors together, and use a LeakyReLU activation layer (negative slope is 0.01) and a Dropout layer to process the aggregated tensor. After that, we apply another 2D convolution layer ($N$ kernels, each kernel size is $[H,k_2]$ and stride is 1) and a Tanh activation to generate the output. 

\subsection{Interactive Convolution Block and Interactive Convolution Network}
The Interactive Convolution (IC) Block is the core module that conducts interactive learning. As shown in Figure~\ref{fig:frwk}(b), we first evenly split the input $S$ into two subsequences, $S_1$ and $S_2$. The splitting scheme can be different. In this study, we split the sequence into odd and even elements. We use two convolution modules to process $S_1$ and $S_2$, transform the outputs of convolution modules by exponential operation, and interact to $S_1$ and $S_2$ with the element-wise product. This process can be expressed by:

\begin{align}
    S_1' &= S_1 \odot \exp \left( Conv(S_2) \right)\\
    S_2' &= S_2 \odot \exp \left( Conv(S_1) \right)
\end{align}
where $\odot$ is the element-wise product operator, $\exp$ is the exponential operator, and $Conv$ is convolution module processing. 

After that, the outputs $S_1'$ and $S_2'$ are further processed by the other two convolution modules and interact with each other by addition and subtraction. This process can be expressed as:

\begin{align}
    S_1'' &= S_1' + Conv(S_2') \\
    S_2'' &= S_2' - Conv(S_1')
\end{align}
where $Conv$ is convolution module processing.

The output of the IC block is the two processed subsequences $S_1''$ and $S_2''$. Note that the four convolution modules in the IC block have the same architecture but use distinct convolution kernels. Therefore, the extracted features from them would contain both homogeneous and heterogeneous information with enhanced representation capabilities \cite{liu2022scinet}. In addition, the interaction between the two subsequences achieves a larger receptive field \cite{liu2022scinet}, thus enhancing the model's feature extraction capabilities.

The Interactive Convolution (IC) Network is constructed by arranging multiple IC blocks hierarchically in a binary-tree-structured framework. As shown in Figure~\ref{fig:frwk}(c), the two output sequences of an IC block in level $i$ will be processed by its two child IC blocks in level $i+1$. This procedure gradually downsamples the input time sequence, thus enabling effective feature learning of different temporal resolutions. The extracted features will be gradually accumulated through the tree so that the model can capture both short-term and long-term dependencies \cite{liu2022scinet}. The outputs of the IC blocks in the bottom level are concatenated and realigned into a new sequence by reversing the splitting operation. The new sequence is added to the origin input $\textbf{X}_t$ through a residual connection \cite{he2016deep}. At last, a fully connected layer is used to generate the prediction $\textbf{Y}_t$. Note that the number of levels of the tree $L$ could be much smaller than  $\log _2T$, given the look-back window length $T$, and $T \mod {2^{L}}=0$ should be satisfied due to the downsampling operation \cite{liu2022scinet}.

\subsection{Loss Function}
We use $L1$ loss in the ICN model training process. The loss function is given by:
\begin{equation}
    Loss = \frac{1}{M}\sum_{i=0}^M \Vert \hat{X}_i - X_i \Vert _1
\end{equation}
where $M$ is the forecast horizon, $\hat{X}_i$ is the predicted demand vector of all areas at time $i$, and $X_i$ is the ground-truth demand.

\section{Case Study}
\label{s:4}
To evaluate the performance of the proposed ICN model, we carried out case studies for the bike-sharing system in Chicago, IL, and the dockless scooter-sharing system in Austin, TX. The performance of the ICN model was compared with the state-of-the-art benchmark models in the two case studies.

\subsection{Data collection and description}
The data used in this study includes the shared micromobility trip data, the point of interest (POI) count data, the demographic data, the transportation supply data, and the weather data in the two case study sites. Table~\ref{tab:variable} presents the descriptive statistics of the input variables used in the two case studies.

\begin{table}[!t]
\caption{Descriptive statistics of input variables} 
\label{tab:variable}
\begin{adjustbox}{width=\textwidth,center}
\begin{tabular}{lllllc}
\hline
\multirow{2}{*}{Variables}                                                                     & \multicolumn{2}{l}{Chicago, IL} & \multicolumn{2}{l}{Austin, TX} & \multirow{2}{*}{Category}                                                                                   \\
                                                                                               & Mean              & Std.             & Mean           & Std.          &                                                                                                          \\ \hline
No. of education facilities                                                                 & 1.39              & 1.31             & 0.66           & 1.06          & \multirow{9}{*}{\begin{tabular}[c]{@{}c@{}}Functional \end{tabular}}               \\
No. of recreational facilities                                                              & 5.57              & 9.20             & 0.16           & 0.42          &                                                                                                          \\
No. of government facilities                                                                & 14.72             & 30.13            & 5.28           & 22.58         &                                                                                                          \\
No. of medical facilities                                                                   & 0.40              & 0.94             & 0.78           & 1.59          &                                                                                                          \\
No. of auto service facilities                                                        & 0.50              & 1.15             & 0.88           & 1.68          &                                                                                                          \\
No. of financial service facilities                                                         & 0.31              & 0.84             & 1.30           & 3.14          &                                                                                                          \\
No. of tourism attractions                                                                  & 0.08              & 0.41             & 0.22           & 0.78          &                                                                                                          \\
No. of hotels                                                                               & 0.09              & 0.48             & 0.32           & 1.31          &                                                                                                          \\
No. of grocery stores                                                                       & 0.41              & 0.72             & 0.59           & 0.91          &                                                                                                          \\ \hline
Population density (per sq. mile)                                                              & 19,094            & 16,231           & 4,798          & 3,937         & \multirow{9}{*}{\begin{tabular}[c]{@{}c@{}}Demographic \end{tabular}}              \\
Pct. of the young population                                                             & 44\%              & 9\%             & 45\%           & 12\%          &                                                                                                          \\
Pct. of the white population                                                             & 42\%              & 30\%             & 74\%           & 15\%          &                                                                                                          \\
Female proportion                                                                              & 51\%              & 5\%              & 49\%           & 6\%           &                                                                                                          \\
\begin{tabular}[c]{@{}l@{}}Pct. of population with BA's \\ degree and above\end{tabular} & 39\%              & 26\%             & 51\%           & 22\%          &                                                                                                          \\
Median household income (USD)                                                            & 68,433            & 37,995           & 80,644         & 37,516        &                                                                                                          \\
Pct. of households own cars                                                              & 65\%              & 18\%             & 91\%           & 8\%           &                                                                                                          \\
Employment density (per mi$^2$)                                                              & 9,807            & 10,052           & 2,748          & 2,043         &                                                                                                          \\ \hline
Bike lane density (mi / mi$^2$)                                                          & 4.26             & 7.13            & 6.37           & 6.46          & \multirow{5}{*}{\begin{tabular}[c]{@{}c@{}}Transportation \\ supply \end{tabular}} \\
WalkScore                                                                                      & 73.13             & 16.93            & 33.55          & 27.82         &                                                                                                          \\
Transit stop density (per mi$^2$)                                                            & 55.77            & 31.60           & 12.77          & 14.41         &                                                                                                          \\
Parking lot density (per mi$^2$)                                                             & 27.69            & 29.33           & 0.86           & 11.72         &                                                                                                          \\
Road network density (mi / mi$^2$)                                                       & 95.04             & 70.66            & 14.44          & 12.20         &                                                                                                          \\ \hline
Daily precipitation (inch)                                                                     & 0.09              & 0.24             & 0.13           & 0.46          & \multirow{4}{*}{Weather}                                                                                       \\
Average temperature (°F)                                                                       & 51.39             & 20.70            & 66.18          & 13.50         &                                                                                                          \\
Snow (inch)                                                                       & 0.10              & 0.50             & -           & -          &                                                                                                          \\
Average wind speed (m/s)                                                                       & -              & -             & 6.44           & 2.71          &                                                                                                          \\ \hline

\end{tabular}
\end{adjustbox}
\end{table}

\begin{figure}[!t]
     \centering
     \begin{subfigure}[h]{0.41\textwidth}
         \centering
         \includegraphics[width=\textwidth]{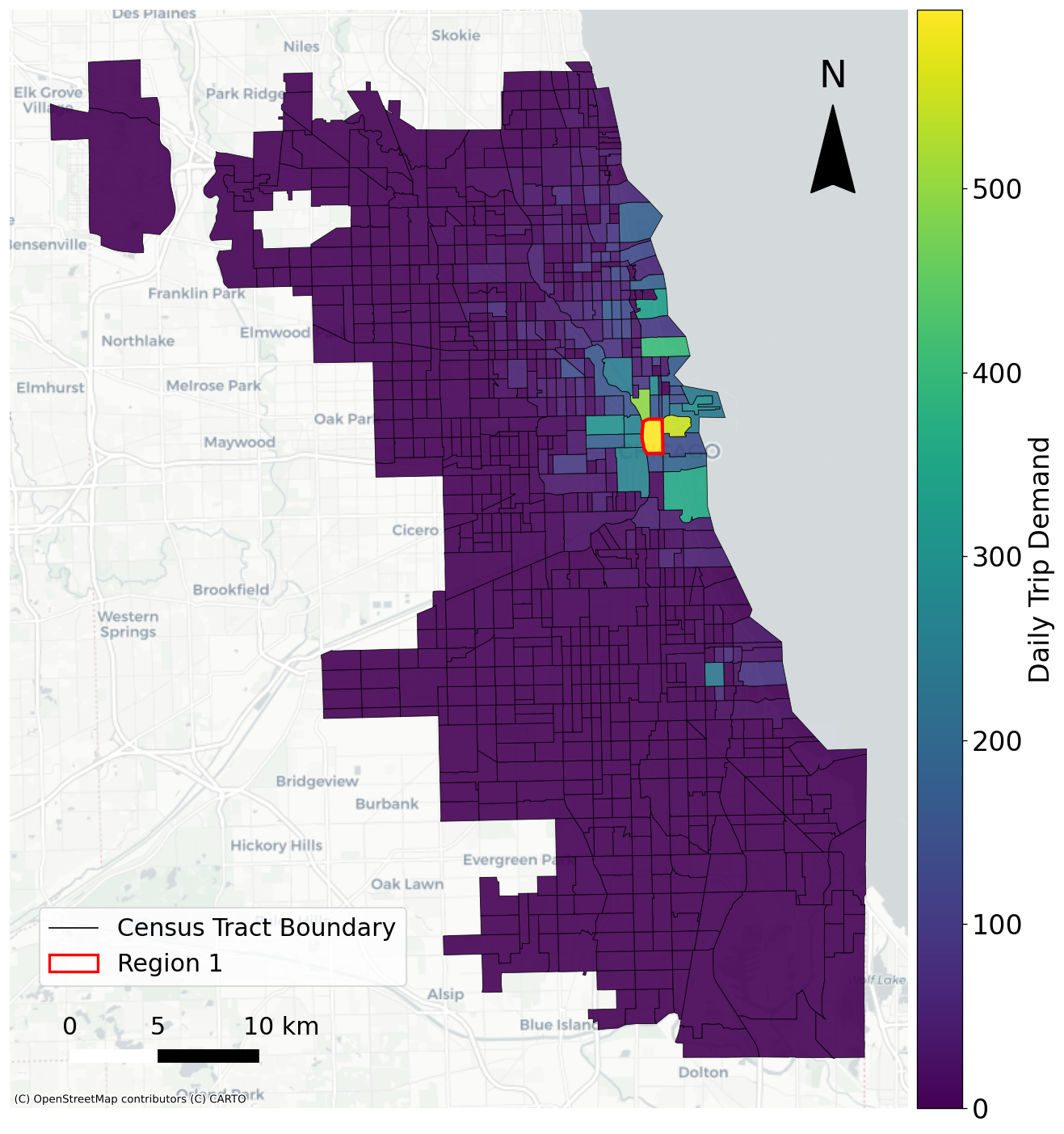}
         \caption{Chicago, IL.}
         \label{fig:dm_chicago}
     \end{subfigure}
     \hfill
     \begin{subfigure}[h]{0.55\textwidth}
         \centering
         \includegraphics[width=\textwidth]{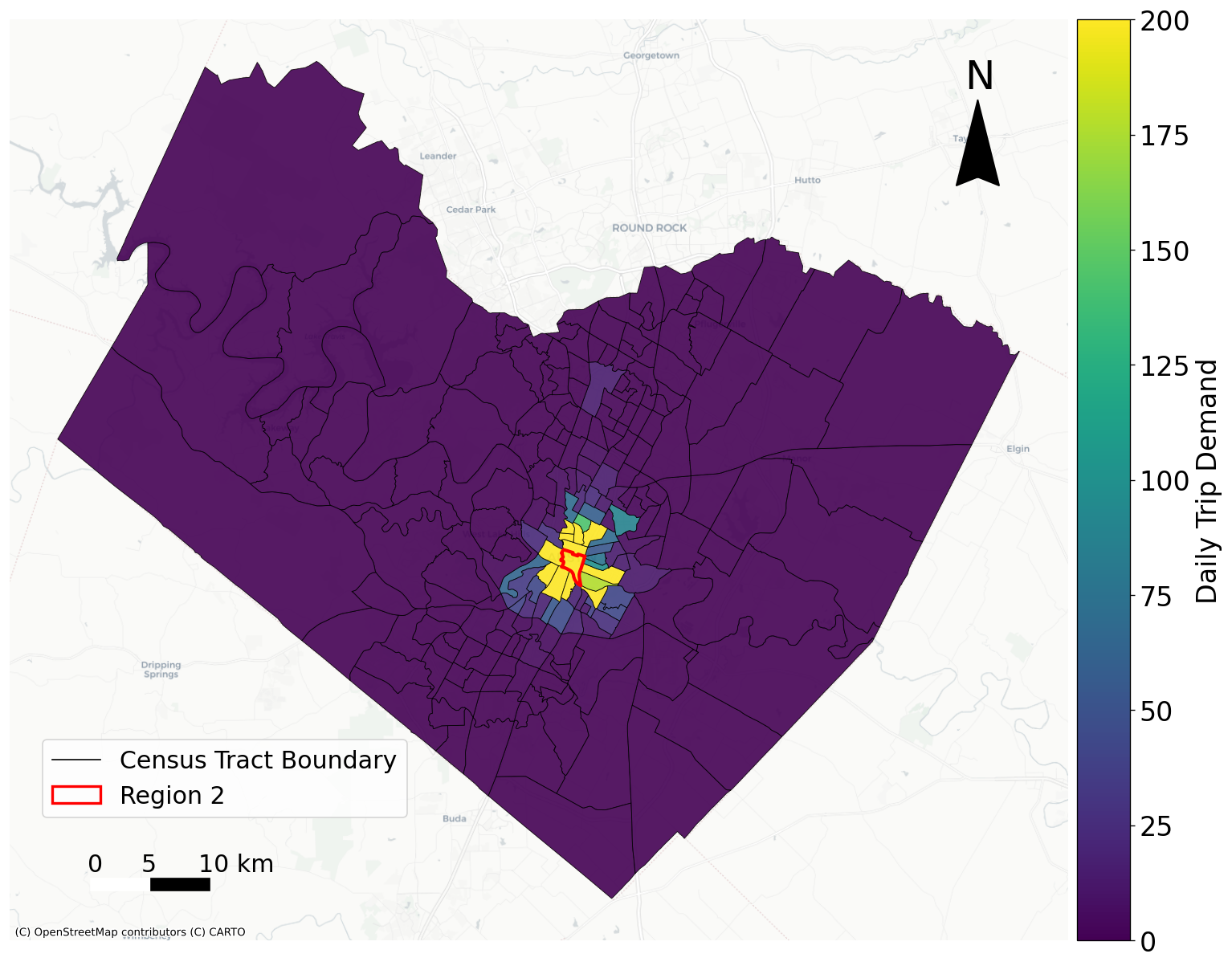}
         \caption{Austin TX.}
         \label{fig:dm_austin}
     \end{subfigure}
        \caption{Spatial distributions of bike-sharing demand in Chicago, IL and dockless scooter-sharing demand in Austin, TX}
        \label{fig:dm}
\end{figure}

The dockless scooter-sharing trip data in Austin, TX were collected from the data-sharing web portal \footnote{https://data.austintexas.gov/Transportation-and-Mobility/Shared-Micromobility-Vehicle-Trips/7d8e-dm7r} operated by the local government. We collected the dockless scooter-sharing trip data from January 1, 2019 to June 30, 2019. The bike-sharing trip data in Chicago, IL were collected from the Divvy Data portal \footnote{https://divvybikes.com/system-data}. The collected data include Divvy bike-sharing trips in Chicago from January 1, 2022 to December 31, 2022. For both case studies, 60\% of the data were used as training set, 20\% of the data were used as validation set, and the remaining 20\% data were used as test set. The spatial distributions of bike-sharing demand in Chicago, IL and dockless scooter-sharing demand in Austin, TX are presented in Figure~\ref{fig:dm}. Given the minimal forecasting needs for areas with exceptionally low demand, our two case studies only incorporated areas with an hourly demand exceeding 1 (including 141 areas in Chicago, IL, and 211 areas in Austin, TX). Note that we compared the model prediction and the observed demand of two regions (i.e., Region 1 of Chicago, IL, and Region 2 of Austin, TX) in Section~\ref{sec:curve}. These regions are also highlighted in Figure~\ref{fig:dm}.

We collected the POI location data from Austin open data portal \footnote{https://data.austintexas.gov}, Chicago data portal \footnote{https://data.cityofchicago.org}, and the Google Maps API \footnote{https://developers.google.com/maps}. The collected POIs included education facilities, recreational facilities, government facilities, medical facilities, automobile service facilities, financial service facilities, tourism attractions, hotels, and grocery stores. We counted the POIs of each category in each block group to aggregate the POIs into census-tract-level. The demographic data included population density, proportion of the young population, proportion of the white population, female proportion, proportion of population with bachelor’s degree and above, median household income, proportion of households that own cars, and employment density. We also collected some transportation supply data, including bike lane density, WalkScore (an index to evaluate the quality of walking environment), transit stop density, parking lot density, and road network density. These variables are highly correlated to the shared micromobility demand \cite{bai2020dockless,merlin2021segment}. The demographic data were collected from the \textit{American Community Survey 5-year} estimates data. We used the \textit{Walkscore.com} API to obtain the WalkScore of a block group centroid and applied geographic information system (GIS) techniques to calculate bike lane density, transit stop density, parking lot density, and road network density. The weather data were collected from the \textit{Global Historical Climatology Network (GHCN)}\footnote{https://www.ncdc.noaa.gov/data-access/land-based-station-data/land-based-datasets/global-historical-climatology-network-ghcn} database. Considering the climate characteristic differences in Chicago and Austin, we use different weather features for the two cases. Daily precipitation, average temperature, and snow are used in the Chicago case, and daily precipitation, average temperature, and average wind speed are used in the Austin case.

\subsection{Model Setting}
The case studies were conducted using an NVIDIA A100 GPU. The kernel size $k_1$ was set to 5 and $k_2$ was set to 3. The hidden size scale $h$ was set to 0.5. The dropout rate was set to 0.5. The number of levels of IC network $L$ was set to 2. We used 60\% of the data for training, 20\% for validation, and 20\% for test. The model was trained using a RMSprop optimizer with $L1$ loss and the initial learning rate was set to 0.001.

\subsection{Model performance}
We evaluated the model performance using Mean Absolute Error (MAE), Root Mean Square Error (RMSE), and Mean Absolute Percentage Error (MAPE). The three evaluation metrics are calculated by:
\begin{equation}
    RMSE=\sqrt{\frac{1}{n}\sum_{i=1}^{n}(\hat{y}_i-y_i)^2}
\end{equation}
\begin{equation}
    MAE=\frac{1}{n}\sum_{i=1}^{n}\vert\hat{y}_i-y_i\vert
\end{equation}
\begin{equation}
    MAPE = \frac{1}{n}\sum_{i=1}^{n}\frac{\vert \hat{y}_i - y_i \vert}{y_i}
\end{equation}\\
\noindent where $y_i$ is the $i$th ground truth value of demand, and $\hat{y}_i$ is the $i$th predicted value of demand. Note that MAPE can be greatly affected by the small values. We only calculate MAPE for samples with demand (ground truth) no less than 10, named MAPE$_{10}$ \cite{tang2021multi}.

We evaluated the proposed ICN model using different input and output settings (i.e., input window and output horizon). The model performance is presented in Table~\ref{tab:icn_performance}. The results show that for both two case studies, the model performed better with a larger input window and a smaller output horizon. This conclusion is consistent with other time series modeling studies: the predictions were more accurate with longer input time sequences, and long-term time series forecasting was more difficult than short-term forecasting \cite{liu2022scinet, zhao2019t, yu2018spatio}. This is because we are able to extract more information from longer input time sequences, and the uncertainty increases when we generate predictions for a more distant future.

\begin{table}[H]
\caption{Performance of the ICN model with different input and output settings. The best results for each output horizon setup are in \textbf{bold}.}
\label{tab:icn_performance}
\begin{tabular}{c|c|ccc|ccc}
\hline
\multirow{2}{*}{\begin{tabular}[c]{@{}c@{}}Output\\ Horizon\end{tabular}} & \multirow{2}{*}{\begin{tabular}[c]{@{}c@{}}Input\\ Window\end{tabular}} & \multicolumn{3}{c|}{Chicago, IL} & \multicolumn{3}{c}{Austin, TX} \\ \cline{3-8} 
                                                                          &                                                                         & MAE      & MAPE$_{10}$     & RMSE      & MAE       & MAPE$_{10}$     & RMSE     \\ \hline
\multirow{4}{*}{1}                                                                 & 24                                                                               & 1.2645                           & \textbf{0.2302}                              & 2.5231                             & 0.8991                           & 0.2166                              & 7.1218                            \\
                                                                                   & 48                                                                               & 1.2454                           & 0.2382                              & 2.4834                             & 0.8877                           & 0.2182                              & 7.5044                            \\
                                                                                   & 96                                                                               & 1.2305                           & 0.2365                              & 2.4808                             & 0.8757                           & 0.2123                              & 6.7353                            \\
                                                                                   & 192                                                                              & \textbf{1.1839}                           & 0.2394                              & \textbf{2.2828}                             & \textbf{0.7715}                           & \textbf{0.2095}                              & \textbf{5.9784}                            \\ \hline
\multirow{4}{*}{6}                                                                 & 24                                                                               & 1.3880                           & \textbf{0.2378}                              & 2.8865                             & 1.1984                           & 0.2193                              & 10.2181                           \\
                                                                                   & 48                                                                               & 1.3762                           & 0.2404                              & 2.8718                             & 1.1549                           & 0.2165                              & 9.7802                            \\
                                                                                   & 96                                                                               & 1.3413                           & 0.2432                              & 2.7957                             & 1.1318                           & 0.2063                              & 8.8603                            \\
                                                                                   & 192                                                                              & \textbf{1.2878}                           & 0.2420                              & \textbf{2.5606}                             & \textbf{0.9400}                           & \textbf{0.2028}                              & \textbf{7.1429}                            \\ \hline
\multirow{4}{*}{12}                                                                & 24                                                                               & 1.4413                           & \textbf{0.2407}                              & 3.0569                             & 1.2202                           & 0.2163                              & 11.0782                           \\
                                                                                   & 48                                                                               & 1.4424                           & 0.2428                              & 3.0969                             & 1.1871                           & 0.2095                              & 10.8888                           \\
                                                                                   & 96                                                                               & 1.4178                           & 0.2439                              & 3.0545                             & 1.1866                           & \textbf{0.2048}                              & 9.5360                            \\
                                                                                   & 192                                                                              & \textbf{1.3336}                           & 0.2416                              & \textbf{2.7069}                             & \textbf{1.0271}                           & 0.2077                              & \textbf{8.3313}                            \\ \hline
\end{tabular}
\end{table}

\subsection{Models comparison}
We compared the proposed ICN model (with three dilated channels) with several benchmark models. The details of these models are described as follows. Note that the input window is 48, and the output horizon is 1. All the models were fine-tuned.

\begin{itemize}
    \item \textbf{HA}: Historical Average is one of the most fundamental statistical models for time series prediction. HA predicts the demand in a specific time period by averaging historical observations. 
    \item \textbf{ARIMA}: Auto-Regressive Integrated Moving Average is a statistical time series prediction model. ARIMA fits a parametric model based on historical observations to predict future demand. The order of ARIMA was set to (1,0,0) in the case study.
    \item \textbf{SVR}: Support Vector Regression \cite{smola2004tutorial} is a machine learning model that uses the same principle as Support Vector Regression (SVM) but for regression problems. We used the Radial Basis Function kernel here. The cost was set to 1, and the gamma was set to 0.02.
    \item \textbf{GBDT}: Gradient Boosting Decision Tree \cite{friedman2001greedy} is a tree-based ensemble machine learning model. In this case study, the number of trees was set to 2000, the maximum depth was set to 7, and the learning rate was set to 0.05. 
    \item \textbf{RF}: Random Forest \cite{breiman2001random} is another tree-based ensemble machine learning method. In this model, the number of trees was set to 110, and the number of features to consider when looking for the best split was set to 7.
    \item \textbf{MLP}: Multiple Layer Perceptron is a classical feedforward artificial neural network. In the case study, we used an MLP model with 100 neurons in the hidden layer. The activation function was ReLU. The learning rate was set to 0.001.  
    \item \textbf{GRU}: Gated Recurrent Unit \cite{cho2014properties} is a widely used RNN model for time series modeling. The learning rate was set to 0.001, the batch size was set to 64, the number of hidden units was 32.
    \item \textbf{LSTM}: Long Short-Term Memory \cite{hochreiter1997long} is another widely used neural network based on the gating mechanism. The learning rate was 0.001, the batch size was 64, and the number of hidden units was 32. 
    \item \textbf{Transformer}: Transformer is a attention-based deep learning network\cite{vaswani2017attention}. In this model, the learning rate was set to 0.005, the batch size was 36, and the number of blocks was 3. 
    \item \textbf{T-GCN}: Temporal Graph Convolutional Network \cite{zhao2019t} is a spatiotemporal graph convolutional neural network that captures spatial and temporal dependency simultaneously using GCN and GRU. The learning rate was set to 0.001. The batch size was set to 64. The number of hidden units was 32.
    \item \textbf{MC-STGCN}: The Multi-Community Spatio-Temporal Graph Convolutional Network \cite{tang2021multi} is an advanced spatiotemporal graph convolutional neural network for passenger demand prediction. The model was trained using Adam optimizer with learning rate of 0.001. The batch size was set to 64. The number of hidden units was 32.
    \item \textbf{STMGT}: Spatio-Temporal Multi-Graph Transformer \cite{xu2023real} is a graph-based spatiotemporal model for dockless scooter-sharing demand forecasting based on GCN and Transformer. The learning rate was set to 0.005. The batch size was set to 32. The number of Transformer blocks was set to 3.
    \item \textbf{SCINet}: The Sample Convolution and Interaction Network \cite{liu2022scinet} is a hierarchical convolutional neural network that conducts sample convolution and interaction for temporal modeling. The model was trained using RMSprop optimizer with initial learning rate of 0.001. The batch size was 8. The hidden size scale was 0.0625 and the dropout rate was 0. The number of levels was 2 and the number of stacks was 1.
   
\end{itemize}

The performance of these models is shown in Table~\ref{tab:comparison}. We can see that the ICN model outperformed all the benchmark models. RNN-based models (i.e., GRU and LSTM) had better prediction performance than statistical models (i.e., HA and ARIMA), classical machine learning models (i.e., SVR, GBDT, and RF), and classical neural network models (i.e., MLP). Among two RNN models, GRU slightly outperformed LSTM. The spatiotemporal neural network models (i.e., T-GCN, ICN) significantly outperformed RNN-based models. This result suggested that learning spatial information improves the prediction accuracy of the model. The performance of SCINet model and ICN model was better compared with other benchmarks. This result suggested that the downsampling process and interactive learning can improve the prediction accuracy, which is consistent with \citep{liu2022scinet}. The ICN model performed better than the SCINet, which indicated that incorporating spatial information (i.e., demographics, functionality, and transportation supply) by channel dilation can further enhance the model performance.

\begin{table}[H]
\caption{Performance of the ICN model and the benchmark models. The look-back window is 48, and the horizon is 1. The best results are in \textbf{bold}.} 
\label{tab:comparison}
\begin{center}
\begin{tabular}{l|ccc|ccc}
\hline
\multirow{2}{*}{Methods} & \multicolumn{3}{c|}{Chicago, IL}            & \multicolumn{3}{c}{Austin, TX}                     \\ \cline{2-7} 
                         & MAE             & MAPE$_{10}$            & RMSE            & MAE             & MAPE$_{10}$            & RMSE            \\ \hline
HA                       & 5.6066          & 0.6563          & 21.422          & 3.3338          & 0.4887          & 6.7861          \\
ARIMA                    & 11.297         & 0.7879          & 28.861          & 3.8302          & 0.7567          & 5.3152          \\
SVR                      & 5.6556          & 0.3580           & 20.925          & 3.3412          & 0.7206          & 7.9320           \\
GBDT                     & 3.7242          & 0.3814          & 20.573          & 1.8502          & 0.3368          & 3.6307          \\
RF                       & 5.4507          & 0.3316          & 20.743          & 1.8911          & 0.3219          & 3.5052          \\
MLP                      & 5.8058          & 0.3972          & 20.758          & 2.0443          & 0.3458          & 3.8477          \\
GRU                      & 2.7542          & 0.3332          & 12.452          & 1.6033          & 0.3622          & 3.3508          \\
LSTM                     & 2.9563          & 0.3351          & 13.110           & 1.6238          & 0.3610           & 3.3940           \\
Transformer              & 2.5954          & 0.3310           & 11.967          & 1.5701          & 0.3265          & 3.2659          \\
T-GCN                    & 2.1826          & 0.3190           & 10.445          & 1.4908          & 0.3223          & 3.0997          \\
MC-STGCN                 & 1.8217          & 0.3001          & 10.370           & 1.4555          & 0.3054          & 2.8897          \\
STMGT                    & 1.0480           & 0.2416          & 7.6636          & 1.3231          & 0.2566          & 2.6721          \\
SCINet                   & 0.9487          & 0.2357          & 8.0568          & 1.3023          & 0.2598          & 2.8765          \\
ICN                      & \textbf{0.8877} & \textbf{0.2182} & \textbf{7.5044} & \textbf{1.2454} & \textbf{0.2382} & \textbf{2.4834} \\ \hline
\end{tabular}
\end{center}
\end{table}

\subsection{Ablation study}

We further conducted an ablation study for the proposed model. We ablated a specific dilated channel or the weather information and then examined the performance of the model to see the contribution of the ablated components. In this ablation study, we generated four ablated models by removing the demographic channel, the functionality channel, the transportation supply channel, or the weather information, respectively. The results are presented in Table~\ref{tab:ablation}. Note that the input window (i.e., look-back window) is fixed to 48. 

We can see that all the ablated models performed worse than the complete model, which means all the ablated components have contributed to the prediction accuracy in the ICN model. For the Chicago case, we observed the greatest increase in prediction errors (i.e., MAE and RMSE) when we ablated weather information from the model, which suggested that weather information was the most important feature in this case. In the Chicago case, when we ablated weather information, the increase in prediction errors was more significant in long-term predictions compared with short-term predictions (2.37\% and 0.66\% increases in MAE and RMSE for the horizon=1 setup, and 9.68\% and 10.02\% increases in MAE and RMSE for the horizon=12 setup), which means that weather information played a more important role in long-term forecasting in this case. For the Austin case, we found the greatest increase in short-term (horizon is 1), medium-term (horizon is 6), and long-term (horizon is 12) MAE when we ablated functionality channel, demographic channel, and transportation supply channel, respectively. Besides, the RMSE increased the most when we ablated weather information, which indicates that weather information contributed the most to predicting extreme values in this case.

\begin{table}[H]
\caption{Performance of ICN model with different channel ablated. The look-back window is 48.} 
\label{tab:ablation}
\begin{adjustbox}{width=0.95\textwidth,center}
\begin{tabular}{c|c|ccc|ccc}
\hline
\multirow{2}{*}{Model}                                                                   & \multirow{2}{*}{\begin{tabular}[c]{@{}c@{}}Output \\ Horizon\end{tabular}} & \multicolumn{3}{c|}{Chicago, IL} & \multicolumn{3}{c}{Austin, TX} \\ \cline{3-8} 
                                                                                         &                                                                            & MAE       & MAPE$_{10}$      & RMSE     & MAE      & MAPE$_{10}$     & RMSE     \\ \hline
\multirow{3}{*}{ICN}                                                                     & 1                                                                          & 1.2454    & 0.2382    & 2.4834   & 0.8877   & 0.2182   & 7.5044   \\
                                                                                         & 6                                                                          & 1.3762    & 0.2404    & 2.8718   & 1.1549   & 0.2165   & 9.7802   \\
                                                                                         & 12                                                                         & 1.4424    & 0.2428    & 3.0969   & 1.1871   & 0.2095   & 10.8888  \\ \hline
\multirow{3}{*}{\begin{tabular}[c]{@{}c@{}}w/o \\ demographic\end{tabular}}              & 1                                                                          & 1.2562    & 0.2374    & 2.4907   & 0.9503   & 0.2209   & 7.5366   \\
                                                                                         & 6                                                                          & 1.3891    & 0.2449    & 2.8870   & 1.2776   & 0.2288   & 10.3432  \\
                                                                                         & 12                                                                         & 1.4996    & 0.2490    & 3.1564   & 1.2902   & 0.2227   & 10.9474  \\ \hline
\multirow{3}{*}{\begin{tabular}[c]{@{}c@{}}w/o \\ functionality\end{tabular}}            & 1                                                                          & 1.2514    & 0.2365    & 2.4880   & 0.9712   & 0.2210   & 7.6323   \\
                                                                                         & 6                                                                          & 1.3742    & 0.2412    & 2.8805   & 1.2189   & 0.2136   & 10.3247  \\
                                                                                         & 12                                                                         & 1.4499    & 0.2471    & 3.1025   & 1.2431   & 0.2221   & 10.9955  \\ \hline
\multirow{3}{*}{\begin{tabular}[c]{@{}c@{}}w/o \\ transportation \\ supply\end{tabular}} & 1                                                                          & 1.2471    & 0.2366    & 2.4887   & 0.9328   & 0.2199   & 7.5445   \\
                                                                                         & 6                                                                          & 1.3827    & 0.2455    & 2.8840   & 1.2206   & 0.2119   & 10.3598  \\
                                                                                         & 12                                                                         & 1.4635    & 0.2456    & 3.1094   & 1.3279   & 0.2225   & 11.1819  \\ \hline
\multirow{3}{*}{\begin{tabular}[c]{@{}c@{}}w/o\\ weather\end{tabular}}                   & 1                                                                          & 1.2749    & 0.2344    & 2.4997   & 0.9344   & 0.2288   & 7.7442   \\
                                                                                         & 6                                                                          & 1.4948    & 0.2488    & 3.1239   & 1.1931   & 0.2124   & 10.4479  \\
                                                                                         & 12                                                                         & 1.5820    & 0.2465    & 3.4072   & 1.2789   & 0.2222   & 11.6142  \\ \hline
\end{tabular}
\end{adjustbox}
\end{table}

\subsection{Prediction results}
\label{sec:curve}
\begin{figure}[H]
     \centering
     \begin{subfigure}[h]{0.80\textwidth}
         \centering
         \includegraphics[width=\textwidth]{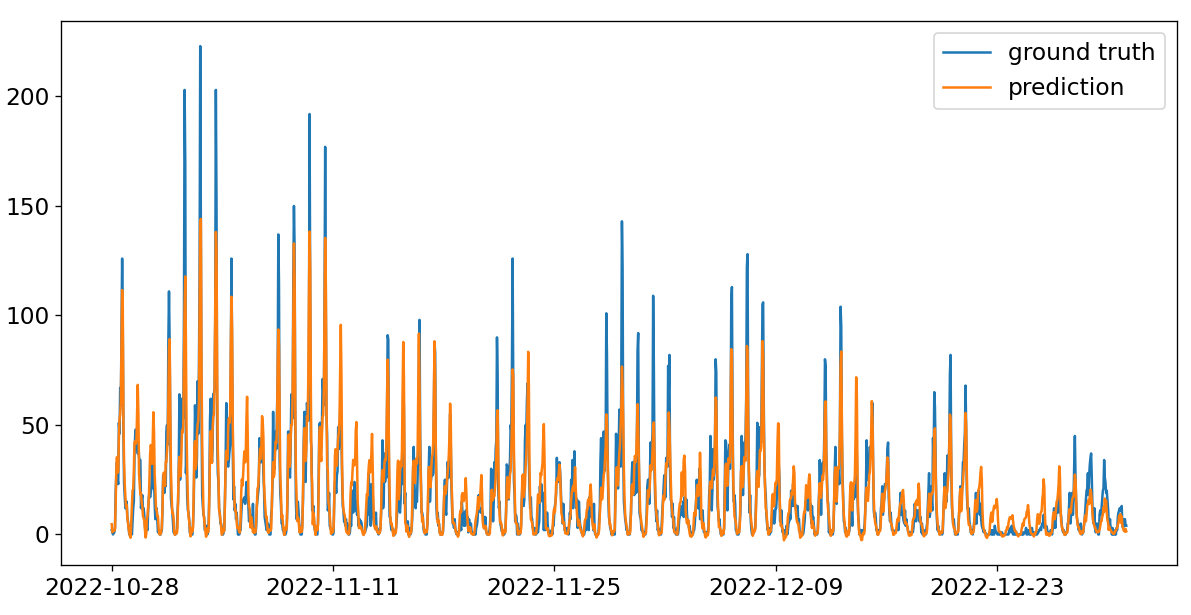}
         \caption{Region 1, Chicago, IL.}
         \label{fig:reg1}
     \end{subfigure}
     \hfill
     \begin{subfigure}[h]{0.81\textwidth}
         \centering
         \includegraphics[width=\textwidth]{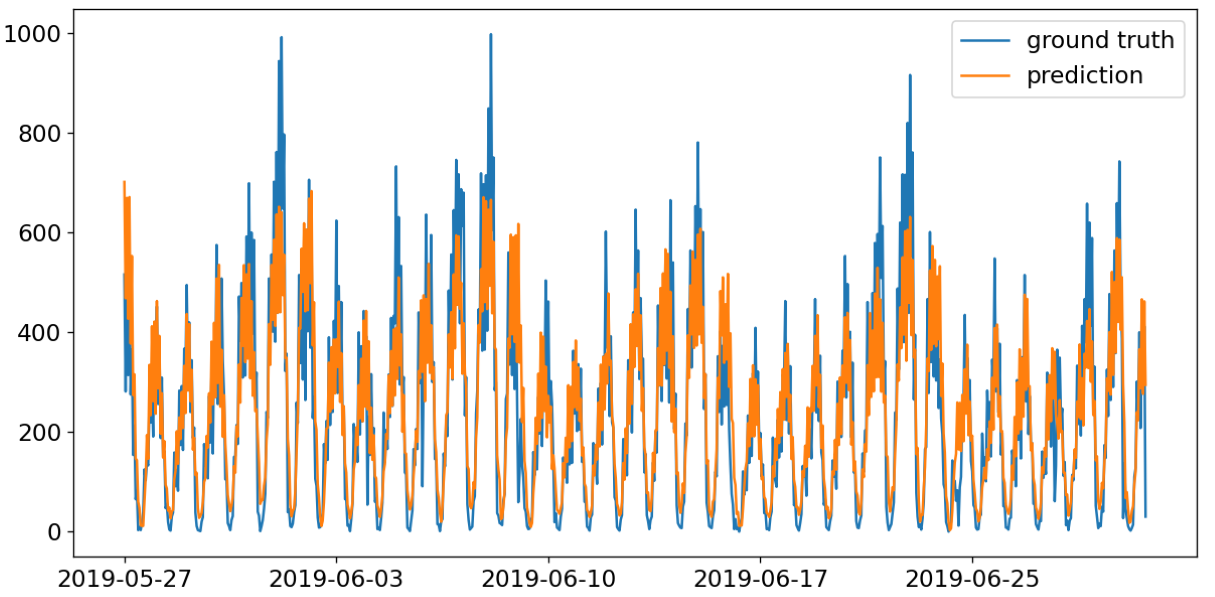}
         \caption{Region 2, Austin TX.}
         \label{fig:reg2}
     \end{subfigure}
        \caption{Comparison of ICN model prediction and ground truth demand.}
        \label{fig:curves}
\end{figure}

We further compared the prediction of the ICN model and the observed demand using test data for two regions with high variance and different demand volumes. The locations of selected regions (i.e., Region 1 of Chicago, IL, and Region 2 of Austin, TX) are highlighted in Figure~\ref{fig:dm}, and the comparison results are presented in Figure~\ref{fig:curves}. 

The figures show the temporal fluctuation of the ground truth and predicted hourly shared micromobility demand in these two areas. The temporal distributions of the observed demand and the model prediction  demonstrate significant temporal coherence. Although the model sometimes cannot fully capture extreme values, the model prediction can effectively follow the temporal fluctuation of observed data in different regions and different time periods. 

\section{Discussion and Conclusion}
\label{s:5}

In this study, we proposed an \textit{Interactive Convolutional Network} (ICN) model for spatiotemporal shared micromobility demand forecasting. The proposed model involved spatial and temporal information for better time series prediction. Specifically, the spatial information (i.e., demographics, functionality, and transportation supply) was selected using travel behavior knowledge. 
The spatial information was embedded in the input demand tensor by the channel dilation process, in which extra input channels were created based on the spatial variables. To simultaneously capture temporal and spatial dependencies, convolution operation was used to process the dilated input tensors. Based on the hierarchical model structure and interactive convolution, the ICN model downsampled the dilated input data and extracted features at different temporal resolutions, then used a fully-connected layer to generate the prediction. 
The proposed methodology of ICN can be readily applied to other transportation applications with higher dimension of channels, such as crash prediction and traffic forecasting. 

The proposed ICN model outperformed all selected benchmark models in two real-world case studies. The outstanding performance of the ICN model resulted from the following reasons. First, the ICN model incorporated multi-dimensional spatial and temporal information adapted from travel behavior knowledge for time series prediction. Since behavioral theories and empirical findings have shown that the shared micromobility travel demand is associated with spatial and temporal factors, such as weather, population density, income, and road network density \cite{bai2020dockless, merlin2021segment, noland2021scootin, xu2023real}, incorporating these factors enhanced the model performance for the travel demand forecasting. Second, we designed a novel channel dilation method to embed spatial information for travel demand forecasting tasks. By creating extra input channels, the proposed channel dilation method allowed us to capture the temporal and spatial dependencies simultaneously using convolution operation, instead of using two separate models to respectively capture temporal and spatial dependencies (e.g., \citep{tang2021multi, xu2023real}). This mechanism improved both predicting accuracy and computational efficiency of the model. Third, we successfully adopted the downsample-convolve-interact process in SCINet \cite{liu2022scinet}, which showcased outstanding performance in several time series prediction tasks (e.g., electricity and traffic speed) but had never been used in travel demand forecasting problems. Using the downsample-convolve-interact process, the model can extract and exchange information at different temporal resolutions, thus learning an effective representation with enhanced predictability \cite{liu2022scinet} and improving the model performance. The model comparison results showed that the model outperformed RNN-based models (e.g., GRU and LSTM) and spatiotemporal models (e.g., T-GCN) even without dilated channels. 

Although the proposed ICN model has achieved high performance in travel demand forecasting, several limitations still require future work. First, we used daily weather condition data to forecast the hourly travel demand in this study. Weather condition data with a smaller time interval (e.g., hourly) may be used to reflect more precise weather conditions and potentially achieve better predictive performance in future work. Second, we assumed that the spatial information is static in this study. In real-world scenarios, spatial information may change at different times of the day. For example, stores and restaurants have specific operating times. Therefore, some functionality variables could be dynamic in the 1-hour resolution. We may incorporate these dynamic characteristics of spatial dependencies in future work.

\section*{Acknowledgements}
This research was supported by the U.S. Department of Transportation through the Southeastern Transportation Research, Innovation, Development and Education (STRIDE) Region 4 University Transportation Center (Grant No. 69A3551747104) and by the University of Florida AI Research Catalyst Fund.

\section*{CRediT Authorship Contribution Statement}
\textbf{Yiming Xu:} Conceptualization; Data curation; Formal analysis; Investigation; Methodology; Software; Validation; Visualization; Writing - original draft. \textbf{Qian Ke:} Conceptualization; Methodology; Supervision; Writing - review \& editing. \textbf{Xiaojian Zhang:} Data curation; Software; Validation; Visualization; Writing - review \& editing. \textbf{Xilei Zhao:} Conceptualization; Funding acquisition; Methodology; Project administration; Resources; Supervision; Writing - review \& editing.

\bibliographystyle{elsarticle-num-names} 
\bibliography{cas-refs}

\begin{thebibliography}{46}
\expandafter\ifx\csname natexlab\endcsname\relax\def\natexlab#1{#1}\fi
\providecommand{\url}[1]{\texttt{#1}}
\providecommand{\href}[2]{#2}
\providecommand{\path}[1]{#1}
\providecommand{\DOIprefix}{doi:}
\providecommand{\ArXivprefix}{arXiv:}
\providecommand{\URLprefix}{URL: }
\providecommand{\Pubmedprefix}{pmid:}
\providecommand{\doi}[1]{\href{http://dx.doi.org/#1}{\path{#1}}}
\providecommand{\Pubmed}[1]{\href{pmid:#1}{\path{#1}}}
\providecommand{\bibinfo}[2]{#2}
\ifx\xfnm\relax \def\xfnm[#1]{\unskip,\space#1}\fi
\bibitem[{Shaheen and Cohen(2021)}]{shaheen2021shared}
\bibinfo{author}{S.~Shaheen}, \bibinfo{author}{A.~Cohen},
\newblock \bibinfo{title}{Shared micromobility: Policy and practices in the
  united states},
\newblock \bibinfo{journal}{A Modern Guide to the Urban Sharing Economy}
  (\bibinfo{year}{2021}) \bibinfo{pages}{166--180}.
\bibitem[{{NACTO}(2022)}]{nacto2022shared}
\bibinfo{author}{{NACTO}},
\newblock \bibinfo{title}{Shared micromobility in the u.s. 2020-2021},
\newblock \bibinfo{journal}{{New York, NY}}  (\bibinfo{year}{2022}).
\bibitem[{McKenzie(2019)}]{mckenzie2019spatiotemporal}
\bibinfo{author}{G.~McKenzie},
\newblock \bibinfo{title}{Spatiotemporal comparative analysis of scooter-share
  and bike-share usage patterns in washington, dc},
\newblock \bibinfo{journal}{Journal of transport geography}
  \bibinfo{volume}{78} (\bibinfo{year}{2019}) \bibinfo{pages}{19--28}.
\bibitem[{Merlin et~al.(2021)Merlin, Yan, Xu, and Zhao}]{merlin2021segment}
\bibinfo{author}{L.~A. Merlin}, \bibinfo{author}{X.~Yan},
  \bibinfo{author}{Y.~Xu}, \bibinfo{author}{X.~Zhao},
\newblock \bibinfo{title}{A segment-level model of shared, electric scooter
  origins and destinations},
\newblock \bibinfo{journal}{Transportation Research Part D: Transport and
  Environment} \bibinfo{volume}{92} (\bibinfo{year}{2021})
  \bibinfo{pages}{102709}.
\bibitem[{Xu et~al.(2022)Xu, Yan, Sisiopiku, Merlin, Xing, and
  Zhao}]{xu2020micromobility}
\bibinfo{author}{Y.~Xu}, \bibinfo{author}{X.~Yan}, \bibinfo{author}{V.~P.
  Sisiopiku}, \bibinfo{author}{L.~A. Merlin}, \bibinfo{author}{F.~Xing},
  \bibinfo{author}{X.~Zhao},
\newblock \bibinfo{title}{Micromobility trip origin and destination inference
  using general bikeshare feed specification data},
\newblock \bibinfo{journal}{Transportation Research Record}
  (\bibinfo{year}{2022}) \bibinfo{pages}{03611981221092005}.
\bibitem[{Schuijbroek et~al.(2017)Schuijbroek, Hampshire, and
  Van~Hoeve}]{schuijbroek2017inventory}
\bibinfo{author}{J.~Schuijbroek}, \bibinfo{author}{R.~C. Hampshire},
  \bibinfo{author}{W.-J. Van~Hoeve},
\newblock \bibinfo{title}{Inventory rebalancing and vehicle routing in bike
  sharing systems},
\newblock \bibinfo{journal}{European Journal of Operational Research}
  \bibinfo{volume}{257} (\bibinfo{year}{2017}) \bibinfo{pages}{992--1004}.
\bibitem[{Jin et~al.(2023)Jin, Wang, Lim, Pan, and Shen}]{jin2023vehicle}
\bibinfo{author}{Z.~Jin}, \bibinfo{author}{Y.~Wang}, \bibinfo{author}{Y.~F.
  Lim}, \bibinfo{author}{K.~Pan}, \bibinfo{author}{Z.-J.~M. Shen},
\newblock \bibinfo{title}{Vehicle rebalancing in a shared micromobility system
  with rider crowdsourcing},
\newblock \bibinfo{journal}{Manufacturing \& Service Operations Management}
  (\bibinfo{year}{2023}).
\bibitem[{Xu et~al.(2023)Xu, Zhao, Zhang, and Paliwal}]{xu2023real}
\bibinfo{author}{Y.~Xu}, \bibinfo{author}{X.~Zhao}, \bibinfo{author}{X.~Zhang},
  \bibinfo{author}{M.~Paliwal},
\newblock \bibinfo{title}{Real-time forecasting of dockless scooter-sharing
  demand: A spatio-temporal multi-graph transformer approach},
\newblock \bibinfo{journal}{IEEE Transactions on Intelligent Transportation
  Systems}  (\bibinfo{year}{2023}).
\bibitem[{Chen et~al.(2020)Chen, Hsieh, Su, Sigalingging, Chen, and
  Leu}]{chen2020predicting}
\bibinfo{author}{P.-C. Chen}, \bibinfo{author}{H.-Y. Hsieh},
  \bibinfo{author}{K.-W. Su}, \bibinfo{author}{X.~K. Sigalingging},
  \bibinfo{author}{Y.-R. Chen}, \bibinfo{author}{J.-S. Leu},
\newblock \bibinfo{title}{Predicting station level demand in a bike-sharing
  system using recurrent neural networks},
\newblock \bibinfo{journal}{IET Intelligent Transport Systems}
  \bibinfo{volume}{14} (\bibinfo{year}{2020}) \bibinfo{pages}{554--561}.
\bibitem[{Lin et~al.(2018)Lin, He, and Peeta}]{lin2018predicting}
\bibinfo{author}{L.~Lin}, \bibinfo{author}{Z.~He}, \bibinfo{author}{S.~Peeta},
\newblock \bibinfo{title}{Predicting station-level hourly demand in a
  large-scale bike-sharing network: A graph convolutional neural network
  approach},
\newblock \bibinfo{journal}{Transportation Research Part C: Emerging
  Technologies} \bibinfo{volume}{97} (\bibinfo{year}{2018})
  \bibinfo{pages}{258--276}.
\bibitem[{Li et~al.(2019)Li, Zhu, Kong, Xu, and Zhao}]{li2019learning}
\bibinfo{author}{Y.~Li}, \bibinfo{author}{Z.~Zhu}, \bibinfo{author}{D.~Kong},
  \bibinfo{author}{M.~Xu}, \bibinfo{author}{Y.~Zhao},
\newblock \bibinfo{title}{Learning heterogeneous spatial-temporal
  representation for bike-sharing demand prediction},
\newblock in: \bibinfo{booktitle}{Proceedings of the AAAI Conference on
  Artificial Intelligence}, volume~\bibinfo{volume}{33}, \bibinfo{year}{2019},
  pp. \bibinfo{pages}{1004--1011}.
\bibitem[{Pan et~al.(2019)Pan, Zheng, Zhang, and Yao}]{pan2019predicting}
\bibinfo{author}{Y.~Pan}, \bibinfo{author}{R.~C. Zheng},
  \bibinfo{author}{J.~Zhang}, \bibinfo{author}{X.~Yao},
\newblock \bibinfo{title}{Predicting bike sharing demand using recurrent neural
  networks},
\newblock \bibinfo{journal}{Procedia computer science} \bibinfo{volume}{147}
  (\bibinfo{year}{2019}) \bibinfo{pages}{562--566}.
\bibitem[{Liu et~al.(2022)Liu, Zeng, Chen, Xu, Lai, Ma, and Xu}]{liu2022scinet}
\bibinfo{author}{M.~Liu}, \bibinfo{author}{A.~Zeng}, \bibinfo{author}{M.~Chen},
  \bibinfo{author}{Z.~Xu}, \bibinfo{author}{Q.~Lai}, \bibinfo{author}{L.~Ma},
  \bibinfo{author}{Q.~Xu},
\newblock \bibinfo{title}{Scinet: time series modeling and forecasting with
  sample convolution and interaction},
\newblock \bibinfo{journal}{Advances in Neural Information Processing Systems}
  \bibinfo{volume}{35} (\bibinfo{year}{2022}) \bibinfo{pages}{5816--5828}.
\bibitem[{Zhao et~al.(2019)Zhao, Song, Zhang, Liu, Wang, Lin, Deng, and
  Li}]{zhao2019t}
\bibinfo{author}{L.~Zhao}, \bibinfo{author}{Y.~Song},
  \bibinfo{author}{C.~Zhang}, \bibinfo{author}{Y.~Liu},
  \bibinfo{author}{P.~Wang}, \bibinfo{author}{T.~Lin},
  \bibinfo{author}{M.~Deng}, \bibinfo{author}{H.~Li},
\newblock \bibinfo{title}{T-gcn: A temporal graph convolutional network for
  traffic prediction},
\newblock \bibinfo{journal}{IEEE Transactions on Intelligent Transportation
  Systems} \bibinfo{volume}{21} (\bibinfo{year}{2019})
  \bibinfo{pages}{3848--3858}.
\bibitem[{Yu et~al.(2018)Yu, Yin, and Zhu}]{yu2018spatio}
\bibinfo{author}{B.~Yu}, \bibinfo{author}{H.~Yin}, \bibinfo{author}{Z.~Zhu},
\newblock \bibinfo{title}{Spatio-temporal graph convolutional networks: a deep
  learning framework for traffic forecasting},
\newblock in: \bibinfo{booktitle}{Proceedings of the 27th International Joint
  Conference on Artificial Intelligence}, \bibinfo{year}{2018}, pp.
  \bibinfo{pages}{3634--3640}.
\bibitem[{Li et~al.(2018)Li, Yu, Shahabi, and Liu}]{li2018diffusion}
\bibinfo{author}{Y.~Li}, \bibinfo{author}{R.~Yu}, \bibinfo{author}{C.~Shahabi},
  \bibinfo{author}{Y.~Liu},
\newblock \bibinfo{title}{Diffusion convolutional recurrent neural network:
  Data-driven traffic forecasting},
\newblock in: \bibinfo{booktitle}{International Conference on Learning
  Representations}, \bibinfo{year}{2018}.
\bibitem[{Guo et~al.(2019)Guo, Lin, Feng, Song, and Wan}]{guo2019attention}
\bibinfo{author}{S.~Guo}, \bibinfo{author}{Y.~Lin}, \bibinfo{author}{N.~Feng},
  \bibinfo{author}{C.~Song}, \bibinfo{author}{H.~Wan},
\newblock \bibinfo{title}{Attention based spatial-temporal graph convolutional
  networks for traffic flow forecasting},
\newblock in: \bibinfo{booktitle}{Proceedings of the AAAI conference on
  artificial intelligence}, volume~\bibinfo{volume}{33}, \bibinfo{year}{2019},
  pp. \bibinfo{pages}{922--929}.
\bibitem[{Wu et~al.(2019)Wu, Pan, Long, Jiang, and Zhang}]{wu2019graph}
\bibinfo{author}{Z.~Wu}, \bibinfo{author}{S.~Pan}, \bibinfo{author}{G.~Long},
  \bibinfo{author}{J.~Jiang}, \bibinfo{author}{C.~Zhang},
\newblock \bibinfo{title}{Graph wavenet for deep spatial-temporal graph
  modeling},
\newblock in: \bibinfo{booktitle}{Proceedings of the 28th International Joint
  Conference on Artificial Intelligence}, \bibinfo{year}{2019}, pp.
  \bibinfo{pages}{1907--1913}.
\bibitem[{Song et~al.(2020)Song, Lin, Guo, and Wan}]{song2020spatial}
\bibinfo{author}{C.~Song}, \bibinfo{author}{Y.~Lin}, \bibinfo{author}{S.~Guo},
  \bibinfo{author}{H.~Wan},
\newblock \bibinfo{title}{Spatial-temporal synchronous graph convolutional
  networks: A new framework for spatial-temporal network data forecasting},
\newblock in: \bibinfo{booktitle}{Proceedings of the AAAI conference on
  artificial intelligence}, volume~\bibinfo{volume}{34}, \bibinfo{year}{2020},
  pp. \bibinfo{pages}{914--921}.
\bibitem[{Bai et~al.(2020)Bai, Yao, Li, Wang, and Wang}]{bai2020adaptive}
\bibinfo{author}{L.~Bai}, \bibinfo{author}{L.~Yao}, \bibinfo{author}{C.~Li},
  \bibinfo{author}{X.~Wang}, \bibinfo{author}{C.~Wang},
\newblock \bibinfo{title}{Adaptive graph convolutional recurrent network for
  traffic forecasting},
\newblock \bibinfo{journal}{Advances in neural information processing systems}
  \bibinfo{volume}{33} (\bibinfo{year}{2020}) \bibinfo{pages}{17804--17815}.
\bibitem[{Jiang et~al.(2023)Jiang, Han, Zhao, and Wang}]{jiang2023pdformer}
\bibinfo{author}{J.~Jiang}, \bibinfo{author}{C.~Han}, \bibinfo{author}{W.~X.
  Zhao}, \bibinfo{author}{J.~Wang},
\newblock \bibinfo{title}{Pdformer: Propagation delay-aware dynamic long-range
  transformer for traffic flow prediction},
\newblock in: \bibinfo{booktitle}{Proceedings of the AAAI conference on
  artificial intelligence}, volume~\bibinfo{volume}{36}, \bibinfo{year}{2023}.
\bibitem[{Geng et~al.(2019)Geng, Li, Wang, Zhang, Yang, Ye, and
  Liu}]{geng2019spatiotemporal}
\bibinfo{author}{X.~Geng}, \bibinfo{author}{Y.~Li}, \bibinfo{author}{L.~Wang},
  \bibinfo{author}{L.~Zhang}, \bibinfo{author}{Q.~Yang},
  \bibinfo{author}{J.~Ye}, \bibinfo{author}{Y.~Liu},
\newblock \bibinfo{title}{Spatiotemporal multi-graph convolution network for
  ride-hailing demand forecasting},
\newblock in: \bibinfo{booktitle}{Proceedings of the AAAI conference on
  artificial intelligence}, volume~\bibinfo{volume}{33}, \bibinfo{year}{2019},
  pp. \bibinfo{pages}{3656--3663}.
\bibitem[{Ke et~al.(2021)Ke, Qin, Yang, Zheng, Zhu, and Ye}]{ke2021predicting}
\bibinfo{author}{J.~Ke}, \bibinfo{author}{X.~Qin}, \bibinfo{author}{H.~Yang},
  \bibinfo{author}{Z.~Zheng}, \bibinfo{author}{Z.~Zhu},
  \bibinfo{author}{J.~Ye},
\newblock \bibinfo{title}{Predicting origin-destination ride-sourcing demand
  with a spatio-temporal encoder-decoder residual multi-graph convolutional
  network},
\newblock \bibinfo{journal}{Transportation Research Part C: Emerging
  Technologies} \bibinfo{volume}{122} (\bibinfo{year}{2021})
  \bibinfo{pages}{102858}.
\bibitem[{Liu et~al.(2019)Liu, Qiu, Li, Wang, Ouyang, and
  Lin}]{liu2019contextualized}
\bibinfo{author}{L.~Liu}, \bibinfo{author}{Z.~Qiu}, \bibinfo{author}{G.~Li},
  \bibinfo{author}{Q.~Wang}, \bibinfo{author}{W.~Ouyang},
  \bibinfo{author}{L.~Lin},
\newblock \bibinfo{title}{Contextualized spatial--temporal network for taxi
  origin-destination demand prediction},
\newblock \bibinfo{journal}{IEEE Transactions on Intelligent Transportation
  Systems} \bibinfo{volume}{20} (\bibinfo{year}{2019})
  \bibinfo{pages}{3875--3887}.
\bibitem[{Bao et~al.(2019)Bao, Liu, and Ukkusuri}]{bao2019spatiotemporal}
\bibinfo{author}{J.~Bao}, \bibinfo{author}{P.~Liu}, \bibinfo{author}{S.~V.
  Ukkusuri},
\newblock \bibinfo{title}{A spatiotemporal deep learning approach for citywide
  short-term crash risk prediction with multi-source data},
\newblock \bibinfo{journal}{Accident Analysis \& Prevention}
  \bibinfo{volume}{122} (\bibinfo{year}{2019}) \bibinfo{pages}{239--254}.
\bibitem[{Basso et~al.(2021)Basso, Pezoa, Varas, and
  Villalobos}]{basso2021deep}
\bibinfo{author}{F.~Basso}, \bibinfo{author}{R.~Pezoa},
  \bibinfo{author}{M.~Varas}, \bibinfo{author}{M.~Villalobos},
\newblock \bibinfo{title}{A deep learning approach for real-time crash
  prediction using vehicle-by-vehicle data},
\newblock \bibinfo{journal}{Accident Analysis \& Prevention}
  \bibinfo{volume}{162} (\bibinfo{year}{2021}) \bibinfo{pages}{106409}.
\bibitem[{Yin et~al.(2021)Yin, Wu, Wei, Shen, Qi, and Yin}]{yin2021deep}
\bibinfo{author}{X.~Yin}, \bibinfo{author}{G.~Wu}, \bibinfo{author}{J.~Wei},
  \bibinfo{author}{Y.~Shen}, \bibinfo{author}{H.~Qi}, \bibinfo{author}{B.~Yin},
\newblock \bibinfo{title}{Deep learning on traffic prediction: Methods,
  analysis, and future directions},
\newblock \bibinfo{journal}{IEEE Transactions on Intelligent Transportation
  Systems} \bibinfo{volume}{23} (\bibinfo{year}{2021})
  \bibinfo{pages}{4927--4943}.
\bibitem[{Zhang et~al.(2017)Zhang, Zheng, and Qi}]{zhang2017deep}
\bibinfo{author}{J.~Zhang}, \bibinfo{author}{Y.~Zheng},
  \bibinfo{author}{D.~Qi},
\newblock \bibinfo{title}{Deep spatio-temporal residual networks for citywide
  crowd flows prediction},
\newblock in: \bibinfo{booktitle}{Proceedings of the AAAI conference on
  artificial intelligence}, volume~\bibinfo{volume}{31}, \bibinfo{year}{2017},
  pp. \bibinfo{pages}{1655--1661}.
\bibitem[{Fu et~al.(2016)Fu, Zhang, and Li}]{fu2016using}
\bibinfo{author}{R.~Fu}, \bibinfo{author}{Z.~Zhang}, \bibinfo{author}{L.~Li},
\newblock \bibinfo{title}{Using lstm and gru neural network methods for traffic
  flow prediction},
\newblock in: \bibinfo{booktitle}{2016 31st Youth Academic Annual Conference of
  Chinese Association of Automation (YAC)}, \bibinfo{organization}{IEEE},
  \bibinfo{year}{2016}, pp. \bibinfo{pages}{324--328}.
\bibitem[{Xu et~al.(2017)Xu, Rahmatizadeh, B{\"o}l{\"o}ni, and
  Turgut}]{xu2017real}
\bibinfo{author}{J.~Xu}, \bibinfo{author}{R.~Rahmatizadeh},
  \bibinfo{author}{L.~B{\"o}l{\"o}ni}, \bibinfo{author}{D.~Turgut},
\newblock \bibinfo{title}{Real-time prediction of taxi demand using recurrent
  neural networks},
\newblock \bibinfo{journal}{IEEE Transactions on Intelligent Transportation
  Systems} \bibinfo{volume}{19} (\bibinfo{year}{2017})
  \bibinfo{pages}{2572--2581}.
\bibitem[{Yao et~al.(2018)Yao, Wu, Ke, Tang, Jia, Lu, Gong, Ye, and
  Li}]{yao2018deep}
\bibinfo{author}{H.~Yao}, \bibinfo{author}{F.~Wu}, \bibinfo{author}{J.~Ke},
  \bibinfo{author}{X.~Tang}, \bibinfo{author}{Y.~Jia}, \bibinfo{author}{S.~Lu},
  \bibinfo{author}{P.~Gong}, \bibinfo{author}{J.~Ye}, \bibinfo{author}{Z.~Li},
\newblock \bibinfo{title}{Deep multi-view spatial-temporal network for taxi
  demand prediction},
\newblock in: \bibinfo{booktitle}{Proceedings of the AAAI Conference on
  Artificial Intelligence}, volume~\bibinfo{volume}{32}, \bibinfo{year}{2018},
  pp. \bibinfo{pages}{2588--2595}.
\bibitem[{Kim et~al.(2019)Kim, Lee, and Sohn}]{kim2019graph}
\bibinfo{author}{T.~S. Kim}, \bibinfo{author}{W.~K. Lee},
  \bibinfo{author}{S.~Y. Sohn},
\newblock \bibinfo{title}{Graph convolutional network approach applied to
  predict hourly bike-sharing demands considering spatial, temporal, and global
  effects},
\newblock \bibinfo{journal}{PloS one} \bibinfo{volume}{14}
  (\bibinfo{year}{2019}) \bibinfo{pages}{e0220782}.
\bibitem[{Li et~al.(2021)Li, Wang, Ren, and Mao}]{li2021intercity}
\bibinfo{author}{H.~Li}, \bibinfo{author}{J.~Wang}, \bibinfo{author}{Y.~Ren},
  \bibinfo{author}{F.~Mao},
\newblock \bibinfo{title}{Intercity online car-hailing travel demand prediction
  via a spatiotemporal transformer method},
\newblock \bibinfo{journal}{Applied Sciences} \bibinfo{volume}{11}
  (\bibinfo{year}{2021}) \bibinfo{pages}{11750}.
\bibitem[{Tang et~al.(2021)Tang, Liang, Liu, Hao, and Wang}]{tang2021multi}
\bibinfo{author}{J.~Tang}, \bibinfo{author}{J.~Liang},
  \bibinfo{author}{F.~Liu}, \bibinfo{author}{J.~Hao},
  \bibinfo{author}{Y.~Wang},
\newblock \bibinfo{title}{Multi-community passenger demand prediction at region
  level based on spatio-temporal graph convolutional network},
\newblock \bibinfo{journal}{Transportation Research Part C: Emerging
  Technologies} \bibinfo{volume}{124} (\bibinfo{year}{2021})
  \bibinfo{pages}{102951}.
\bibitem[{Cai et~al.(2020)Cai, Janowicz, Mai, Yan, and Zhu}]{cai2020traffic}
\bibinfo{author}{L.~Cai}, \bibinfo{author}{K.~Janowicz},
  \bibinfo{author}{G.~Mai}, \bibinfo{author}{B.~Yan}, \bibinfo{author}{R.~Zhu},
\newblock \bibinfo{title}{Traffic transformer: Capturing the continuity and
  periodicity of time series for traffic forecasting},
\newblock \bibinfo{journal}{Transactions in GIS} \bibinfo{volume}{24}
  (\bibinfo{year}{2020}) \bibinfo{pages}{736--755}.
\bibitem[{Xu et~al.(2020)Xu, Dai, Liu, Gao, Lin, Qi, and Xiong}]{xu2020spatial}
\bibinfo{author}{M.~Xu}, \bibinfo{author}{W.~Dai}, \bibinfo{author}{C.~Liu},
  \bibinfo{author}{X.~Gao}, \bibinfo{author}{W.~Lin}, \bibinfo{author}{G.-J.
  Qi}, \bibinfo{author}{H.~Xiong},
\newblock \bibinfo{title}{Spatial-temporal transformer networks for traffic
  flow forecasting},
\newblock \bibinfo{journal}{arXiv preprint arXiv:2001.02908}
  (\bibinfo{year}{2020}).
\bibitem[{Bai and Jiao(2020)}]{bai2020dockless}
\bibinfo{author}{S.~Bai}, \bibinfo{author}{J.~Jiao},
\newblock \bibinfo{title}{Dockless e-scooter usage patterns and urban built
  environments: A comparison study of austin, tx, and minneapolis, mn},
\newblock \bibinfo{journal}{Travel behaviour and society} \bibinfo{volume}{20}
  (\bibinfo{year}{2020}) \bibinfo{pages}{264--272}.
\bibitem[{Xu et~al.(2021)Xu, Yan, Liu, and Zhao}]{xu2021identifying}
\bibinfo{author}{Y.~Xu}, \bibinfo{author}{X.~Yan}, \bibinfo{author}{X.~Liu},
  \bibinfo{author}{X.~Zhao},
\newblock \bibinfo{title}{Identifying key factors associated with ridesplitting
  adoption rate and modeling their nonlinear relationships},
\newblock \bibinfo{journal}{Transportation Research Part A: Policy and
  Practice} \bibinfo{volume}{144} (\bibinfo{year}{2021})
  \bibinfo{pages}{170--188}.
\bibitem[{Noland(2021)}]{noland2021scootin}
\bibinfo{author}{R.~B. Noland},
\newblock \bibinfo{title}{Scootin’in the rain: Does weather affect
  micromobility?},
\newblock \bibinfo{journal}{Transportation Research Part A: Policy and
  Practice} \bibinfo{volume}{149} (\bibinfo{year}{2021})
  \bibinfo{pages}{114--123}.
\bibitem[{He et~al.(2016)He, Zhang, Ren, and Sun}]{he2016deep}
\bibinfo{author}{K.~He}, \bibinfo{author}{X.~Zhang}, \bibinfo{author}{S.~Ren},
  \bibinfo{author}{J.~Sun},
\newblock \bibinfo{title}{Deep residual learning for image recognition},
\newblock in: \bibinfo{booktitle}{Proceedings of the IEEE conference on
  computer vision and pattern recognition}, \bibinfo{year}{2016}, pp.
  \bibinfo{pages}{770--778}.
\bibitem[{Smola and Sch{\"o}lkopf(2004)}]{smola2004tutorial}
\bibinfo{author}{A.~J. Smola}, \bibinfo{author}{B.~Sch{\"o}lkopf},
\newblock \bibinfo{title}{A tutorial on support vector regression},
\newblock \bibinfo{journal}{Statistics and computing} \bibinfo{volume}{14}
  (\bibinfo{year}{2004}) \bibinfo{pages}{199--222}.
\bibitem[{Friedman(2001)}]{friedman2001greedy}
\bibinfo{author}{J.~H. Friedman},
\newblock \bibinfo{title}{Greedy function approximation: a gradient boosting
  machine},
\newblock \bibinfo{journal}{Annals of statistics}  (\bibinfo{year}{2001})
  \bibinfo{pages}{1189--1232}.
\bibitem[{Breiman(2001)}]{breiman2001random}
\bibinfo{author}{L.~Breiman},
\newblock \bibinfo{title}{Random forests},
\newblock \bibinfo{journal}{Machine learning} \bibinfo{volume}{45}
  (\bibinfo{year}{2001}) \bibinfo{pages}{5--32}.
\bibitem[{Cho et~al.(2014)Cho, Van~Merri{\"e}nboer, Bahdanau, and
  Bengio}]{cho2014properties}
\bibinfo{author}{K.~Cho}, \bibinfo{author}{B.~Van~Merri{\"e}nboer},
  \bibinfo{author}{D.~Bahdanau}, \bibinfo{author}{Y.~Bengio},
\newblock \bibinfo{title}{On the properties of neural machine translation:
  Encoder-decoder approaches},
\newblock \bibinfo{journal}{arXiv preprint arXiv:1409.1259}
  (\bibinfo{year}{2014}).
\bibitem[{Hochreiter and Schmidhuber(1997)}]{hochreiter1997long}
\bibinfo{author}{S.~Hochreiter}, \bibinfo{author}{J.~Schmidhuber},
\newblock \bibinfo{title}{Long short-term memory},
\newblock \bibinfo{journal}{Neural computation} \bibinfo{volume}{9}
  (\bibinfo{year}{1997}) \bibinfo{pages}{1735--1780}.
\bibitem[{Vaswani et~al.(2017)Vaswani, Shazeer, Parmar, Uszkoreit, Jones,
  Gomez, Kaiser, and Polosukhin}]{vaswani2017attention}
\bibinfo{author}{A.~Vaswani}, \bibinfo{author}{N.~Shazeer},
  \bibinfo{author}{N.~Parmar}, \bibinfo{author}{J.~Uszkoreit},
  \bibinfo{author}{L.~Jones}, \bibinfo{author}{A.~N. Gomez},
  \bibinfo{author}{{\L}.~Kaiser}, \bibinfo{author}{I.~Polosukhin},
\newblock \bibinfo{title}{Attention is all you need},
\newblock \bibinfo{journal}{Advances in neural information processing systems}
  \bibinfo{volume}{30} (\bibinfo{year}{2017}).

\end{thebibliography}





\end{document}